# Weather conditions at Timau National Observatory from ERA5

R. Priyatikanto[1,2], A.G. Admiranto[1], T. Djamaluddin[1], A. Rachman[1], D.D. Wijaya[3]

[1]Research Centre for Space, National Research and Innovation Agency, Bandung, Indonesia

[2]School of Geography and Natural Sciences, University of Southampton, United Kingdom

[3]Geodesy Research Group, Faculty of Earth Sciences and Technology, Institut Teknologi Bandung, Bandung, Indonesia

**ABSTRACT**

A new observatory site should be investigated for its local climate conditions to see its potential and limitations. In this respect, we examine several meteorological parameters at the site of Timau National Observatory, Indonesia using the ERA5 dataset from 2002 to 2021. Based on this dataset, we conclude that the surface temperature at Timau is around 18.9 $^{o}$C with relatively small temperature variation (1.5$^{o}$C) over the day. This temperature stability is expected to give advantages to the observatory. In terms of humidity and water vapour, Timau is poor for infrared observations as the median precipitable water vapour exceeds 18 mm, even during the dry season. However, near-infrared observations are feasible. Even though our cloud cover analysis confirms the span of the observing season in the region, we find a significant discrepancy between the clear sky fraction derived from the ERA5 dataset and the one estimated using satellite imagery. Aside from the indicated bias, our results provide insights and directions for the operation and future development of the observatory.

Keywords: Observatories (1147), Astronomical site protection (94)

## 1. INTRODUCTION

Sites with darker, tranquil, and more transparent skies are ideal for ground-based astronomical observations. At these sites, the degradation of electromagnetic signals from astronomical objects due to the Earth's atmosphere is minimal. In general, astronomical sites are selected based on the ease of access, remoteness from the source of artificial light pollution, and the local climate characteristics such as cloud cover, atmospheric transparency, humidity, temperature, and precipitation (e.g., Vernin et al. 2011; Aksaker et al. 2020). Thus, it is very important to acquire long-term statistics on essential meteorological parameters prior to the decision on where to place a new observing facility. Continuous monitoring should also be performed so the quality of astronomical observations can be assured and scientific productivity can be maintained. Comprehensive site testing and monitoring give valuable information on the characteristics of the atmosphere and the optimal circumstances for astronomical observations (Tillayev et al. 2023). Even further, a projection of the future climatic conditions of some astronomical sites can be done, using a realistic climate model and scenario, to mitigate the impact of climate change on the expected observing conditions (Haslebacher et al. 2022).

Atmospheric temperature becomes the most crucial meteorological parameter that has been recorded for a very long time (Plaut, Ghil, and Vautard 1995). It is also one key variable for astronomical observations. Spatial and temporal stability of the air temperature affects turbulence in the local atmosphere which in turn determines the quality of astronomical images which is also known as seeing (Coulman 1985). A smaller drop in temperature during a clear night is expected to be a sign of good observing conditions. This became the reason behind the selection of the Chilean Andes as the site for the first European Southern Observatory, opting out the alternatives in South Africa (Blaauw 1989, 1991). Air temperature also has an obvious role in the hydrological cycle and plays a key role in energy transport. Water content in the atmosphere is governed by temperature through some mechanisms including evaporation, condensation, and precipitation. Atmospheric transparency in the infrared and microwave windows anti-correlates with the amount of water vapour in the atmosphere as water molecules intensely absorb electromagnetic radiation at these windows.

In brief, drier is better for astronomy. On the contrary, a tropical region with a humid climate like Indonesia is far from ideal for astronomical observation. However, a study by Hidayat et al. 2012, which was later endorsed by Wang et al. 2022 and Priyatikanto et al. 2023, shows that the southeastern part of Indonesia can be a prospective location for a new observatory in this tropical country. Annually, the percentage of usable nights is about 66%. In the region, the probability of getting a clear sky exceeds 80% during the observing season which lasts from April to October. Among several alternatives, Timau Mountain in Timor Island was selected for a new observatory.

Timau National Observatory (123.947° E, 9.597° S) and accessibility, the Timau site is deemed the best choice for a medium-sized astronomical observatory in the Mutis-Timau region. Even though the climate of Timor Island is influenced by the drier Australian Continent, a mountainous tropical forest with moderately dense trees, shrubs, savannas, and seasonal rivers can be found (Pujiono et al. 2019).

As mentioned by Mumpuni et al. 2018, Timau National Observatory will host a 3.8-m segmented telescope which is a sister of the Seimei Telescope in Okayama, Japan (Kurita et al. 2020). With its boosted agility, the telescope is optimized for rapid response observations. Three-band imager that simultaneously captures images at g, r, and i bands will be the first generation instrument for the telescope, together with a near-infrared camera that operates up to 1.8 microns. Some smaller telescopes will also be operated on-site to survey and monitor space and celestial objects.

To support the operation and planning for future development, we analyze essential meteorological parameters at the Timau site. Long-term in-situ measurement is not available for this purpose so we rely on the latest global climate reanalysis dataset produced by the European Centre for Medium-Range Weather Forecast (ECMWF). This so-called ERA5 (Hersbach et al. 2020) has been utilized for characterizing some astronomical sites. For instance, Shikhovtsev et al. 2021 evaluated astroclimatic conditions at some places in Vietnam using a decade of ERA5 dataset. In line with that, Zhu et al. 2023 utilized reanalysis data from ECMWF to describe the wind profile over Tibet, in relation to the astronomical seeing. ERA5 and other reanalysis datasets also provide temporal cloud cover estimates which is very important for astronomical site evaluation (Ningombam et al. 2021). Precipitable water vapor (PWV) which is a determinant factor of atmospheric transparency in infrared and microwave domains is also available in ERA5. Compatible with the value derived from radiosonde and satellite-based measurements (Huang et al. 2021; Wijaya et al. 2023), ERA5 enables astronomers to see long-term variations of PWV and some other meteorological parameters relevant to astronomical observations (Li et al. 2020; Bolbasova, Shikhovtsev, and Ermakov 2023).

## 2. DATA AND METHOD

We employ a decadal dataset from the fifth generation of the European Centre for Medium-Range Weather Forecasts Reanalysis (ERA5, Hersbach et al. 2020). This dataset is a product of the Integrated Forecasting System cycle 41r2 that incorporates relevant physics and core dynamics in the atmosphere. For bias correction, the assimilation is performed using observational data from around the globe. With global coverage and impressive spatiotemporal resolutions, ERA5 and its predecessor (ERA-Interim) have been used in some essential climatological assessments, such as the annual reviews of the state of the climate by the World Meteorological Organization and the Intergovernmental Panel on Climate Change. The astronomical community also utilizes the dataset to evaluate the weather and climate conditions of some astronomical sites (Li et al. 2020; Huang et al. 2021). ERA5 has a horizontal resolution of 0.25° or 31 km on the equator and captures meteorological parameters at different pressure levels up to 1 Pa (~100 km altitude). Hourly data from 1950 is available to download from the Climate Data Store.

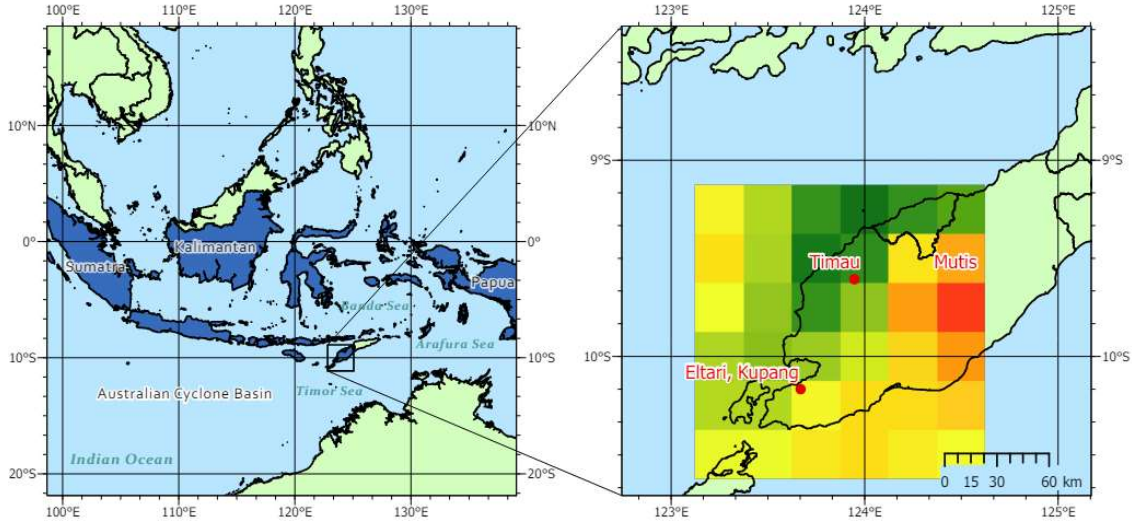

Figure 1. The location of Timau National Observatory in the southeastern part of Indonesia. The map on the right locates Timau and Kupang with overlaid 31-km grid of the ERA5 dataset.

Specific to the current study, we utilize ERA5 hourly data on pressure levels (1 to 1000 hPa) from 2002 to 2021 covering an area of 1.5° × 1.5° around the Timau and Kupang (see Figure 1). For a particular time, the dataset describes meteorological conditions in three-dimensional space of approximately 180 × 180 × 40 km³ above Timau and its surrounding area. It contains cloud cover, temperature, humidity, and wind speed (eastward and northward components) with a spatial resolution of 31 km on the ground. The selected meteorological parameters are directly associated with astronomical observations at optical and near-infrared domains. To obtain the appropriate values at Timau (123.947° E, 9.597° S, 1300 masl), we apply a cubic interpolation scheme based on the regularly-gridded data from ERA5. During the process, three-dimensional interpolation is decomposed into three separate cubic interpolations (along longitude, latitude, and height) and then combined into the final solution. Python xarray package (Hoyer and Hamman 2017) with its interp function helps to perform this multi-dimensional interpolation.

Specific for the cloud cover (CC), we aggregate the cloud cover at low (< 2000 m), medium (2000 – 6000 m), and high altitudes (> 6000 m) using the following rule (see Eq. 1 of Hellemeier and Hickson 2019):

$$CC = \max\left(\left[\frac{2000-a}{2000}CC_{\text{low}}, CC_{\text{medium}}, CC_{\text{high}}\right]\right). \quad (1)$$

Considering the elevation of Timau (a = 1300), we apply simple scaling to the low-altitude cloud cover.

Additionally, we compute the amount of precipitable water vapour (PWV) in order to characterize the atmospheric transparency at the near-infrared window. Following its definition as the integrated amount of water vapour along the column extending from the Earth's surface (isobaric pressure of $p_{surf}$ = 863 hPa) to the top of the atmosphere (isobaric pressure of $p_{top}$ = 1 hPa), PWV (in mm) is computed using the following formula:

$$PWV = \frac{10}{\rho g} \int_{p_{top}}^{p_{surf}} q\, dp, \qquad (2)$$

where q is the specific humidity (dimensionless), which defines the mass of water vapour per kilogram of moist air. The density of water (1000 kg m$^{-3}$) is symbolized as ρ, whereas the Earth's surface gravity is g = 9.8 m s$^{-2}$.

To check the validity of the ERA5 dataset, we use daily aggregate weather data from 2020 to 2021 acquired at the Eltari Weather Station (123.673° E, 10.171° S), which is approximately 50 km from the Timau site. The station is operated by the Indonesian Meteorology Climatology and Geophysics Agency (BMKG) to mainly support aviation activities. As summarized in Figure 2, the daily average of temperature from ERA5 fairly agrees with the on-site measurement as indicated by the coefficient of determination $R^2$ = 0.58 and root mean square deviation RMSD = 1.46 °C. Nevertheless, the average temperature from ERA5 is somewhat lower than the one from BMKG as specified by –1.24 °C bias. Even better scores are achieved for relative humidity ($R^2$ = 0.79, RMSD = 4.99%, bias = 0.47%). The resolution of horizontal wind speed data from BMKG is relatively low, but it is still comparable to ERA5 with $R^2$ = 0.67, RMSD = 1.00 m/s, and bias = –0.45 m/s.

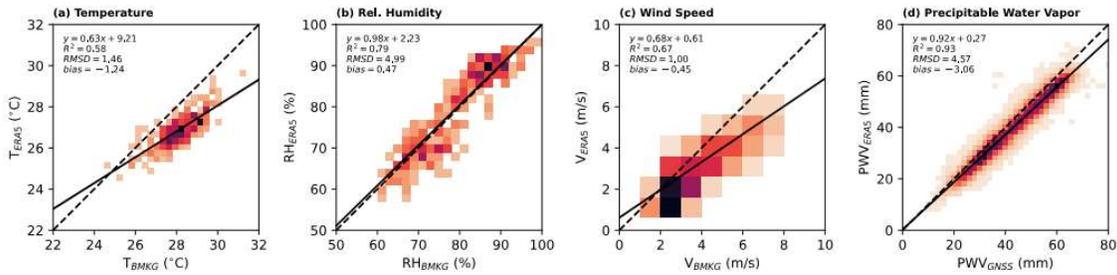

Figure 2. Correlation between weather parameters at Kupang from ERA5 and in-situ measurements. Temperature, relative humidity, and wind speed are from the weather station operated by BMKG. Precipitable water vapor is based on the time delay of GNSS signals (Wijaya et al. 2023). Linear fit to the data, coefficient of determination (R2), root mean square deviation (RMSD), and bias represent the agreement between the datasets.

We also compare PWV from ERA5 to the satellite-based value which is derived from the hydrostatic delay of the Global Navigation Satellite System (GNSS) signals (Wijaya et al. 2023). The GNSS receiver for this data is located at Kupang (123.596° E, 10.169° S). Using hourly data from 2012 to 2021, we demonstrate an excellent agreement between the ERA5 and the GNSS-based PWV. The coefficient of determination is 0.93 while the RMSD and the bias are 4.57 mm and -3.06 mm, respectively.

## 3. RESULTS
### 3.1 Cloud Cover

Timau has a moderate clear sky fraction with obvious variation over time (see Figure 3). During the wettest months (December to March), the median cloud cover (CC) exceeds 0.90 such that astronomical observation at the optical-infrared window becomes impossible. The average CC declines to under 0.50 during the dry season and reaches a minimum value of around 0.20 in August. Consequently, the observing season may last from April to October each year.

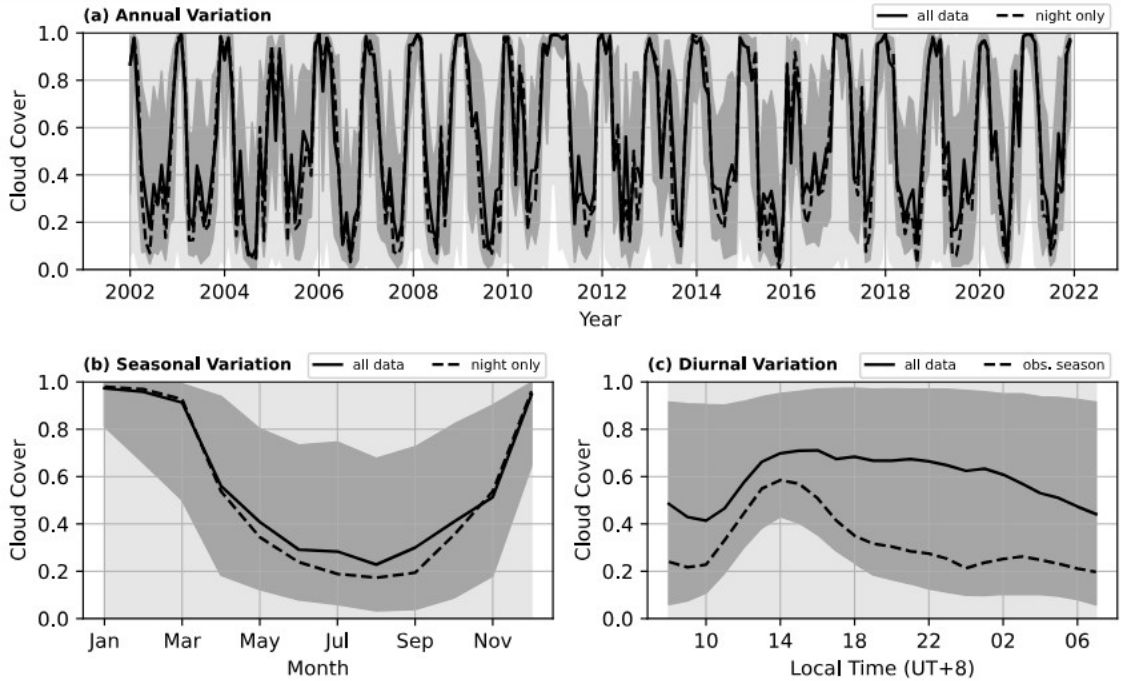

Figure 3. Time series plot showing annual (a), seasonal (b), and diurnal variation (c) of cloud cover at Timau site. The solid line marks the median value of the whole data while the dashed line is for the time-filtered data. The shaded area encompasses the inter-quartile range (percentile 25 to 75) of the temperature whereas the minimum to maximum range is lightly shaded.

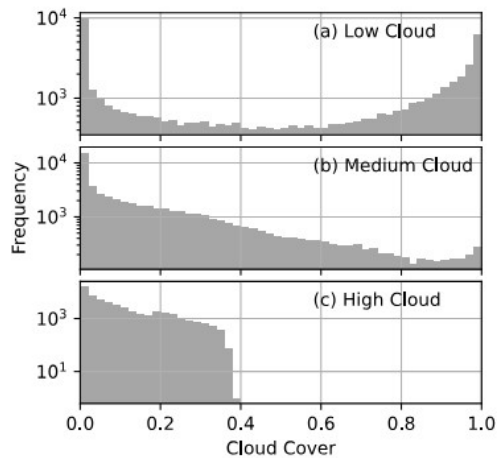

Figure 4. Frequency distribution of cloud cover at different altitude ranges (low: < 2000m, medium: 2000 – 6000 m, high: > 6000 m.

In general, lower CC happens in the morning while the daily maximum is reached at 14.00 local time. However, the daily variation during the observing season has distinctive characteristics compared to the one extracted from the whole dataset. During the observing season, low CC is maintained for a longer duration at night, especially after midnight. This condition gives advantages to astronomical observations on site. In the ERA5 dataset, the cloud cover is described as the proportion of the grid box covered by clouds. This parameter can be dissected into three parameters, namely low-altitude cloud cover (below 2000 m), mid-altitude cloud cover (between 2000 and 6000 m), and high-altitude cloud cover (above 6000 m). As indicated in Figure 4, for altitudes below 6000 m, the distribution function of CC is more or less exponential, i.e., the chance of getting higher CC decreases exponentially. This does not apply to the high-altitude cloud cover whose distribution shows a high rise on the right end. The appearance of cirrus clouds, which is more frequent in tropical regions, is associated with overcast conditions at Timau.

3.2 Temperature

Hot and humid are two main characteristics of the climate in the tropical region, including Timau. The average surface temperature at this site is approximately 19 ºC with monthly medians ranging from 16 to 20 ºC. Statistically, the majority of the data vary within 16.5 (percentile-5) and 20.5 ºC (percentile-95). Figure 5 summarizes the variability of the surface temperature, aggregated at different temporal domains. From the annual variation, there is a slight increase in temperature during the period from 2002 until 2022. Assuming a linear trend, the warming rate of 0.03 ºC per year is observed. Seasonal variation is clearly seen in the plot where the pattern starts with a mild temperature followed by a small bump in April and a shallow drop in August. The monthly median temperature reaches the maximum in October. Nevertheless, some anomalous years with higher annual means and warmer dry seasons are identified, i.e. 2010, 2016-2017, and 2021. These anomalies are associated with the warmer phase of the El Niño Southern Oscillation (ENSO).

The lower-left panel of Figure 5 clarifies the seasonal variation in terms of surface temperature. During the wet season, which occurs from October to April, the temperature variation is smaller than the variation during the dry season. The interquartile range (IQR) of temperature in the wet season is around 0.5 ºC, while its value in the dry season reaches 0.8 ºC. This pattern is in accordance with the fact that humid air retains heat better than dry air so that temperature variation is suppressed during the wet season.

Based on the aggregated data, the diurnal variation of the temperature at Timau is considerably small. During the day (06.00 - 18.00 local time), the median values range from 18.8 to 19.1 ºC, while the nighttime medians vary from 18.7 to 18.9 ºC. The amplitude of the variation is enhanced during the observing season that is coincident with the dry season (dashed line in the lower-right panel of Figure 5). During this season, the median temperature drops by ~ 0.8 ºC approximately four hours after reaching its peak at 14.00 local time.

The diurnal stability of the surface temperature at Timau benefits astronomical observation. Smaller variation in temperature leads to suppressed atmospheric turbulence and better seeing condition near the ground. Additionally, a shallower drop in temperature during dusk means a shorter time is required for acclimatization to equate the temperature inside and outside the telescope enclosure. Moreover, the urge to operate an air conditioner to lower the temperature inside the telescope enclosure is suppressed.

To further confirm the finding above, we also computed the minimum and maximum temperatures on a daily basis and summarized their statistics. The global median of the minimum temperature is 18.2 ºC, whereas the maximum is 19.7 ºC. In the nighttime, the typical values of the minimum and maximum temperatures, respectively, are 18.3 and 19.3 ºC. Additionally, the temporal change of the temperature ($|\Delta T/\Delta t|$) is around the median of 0.15 ºC/hour.

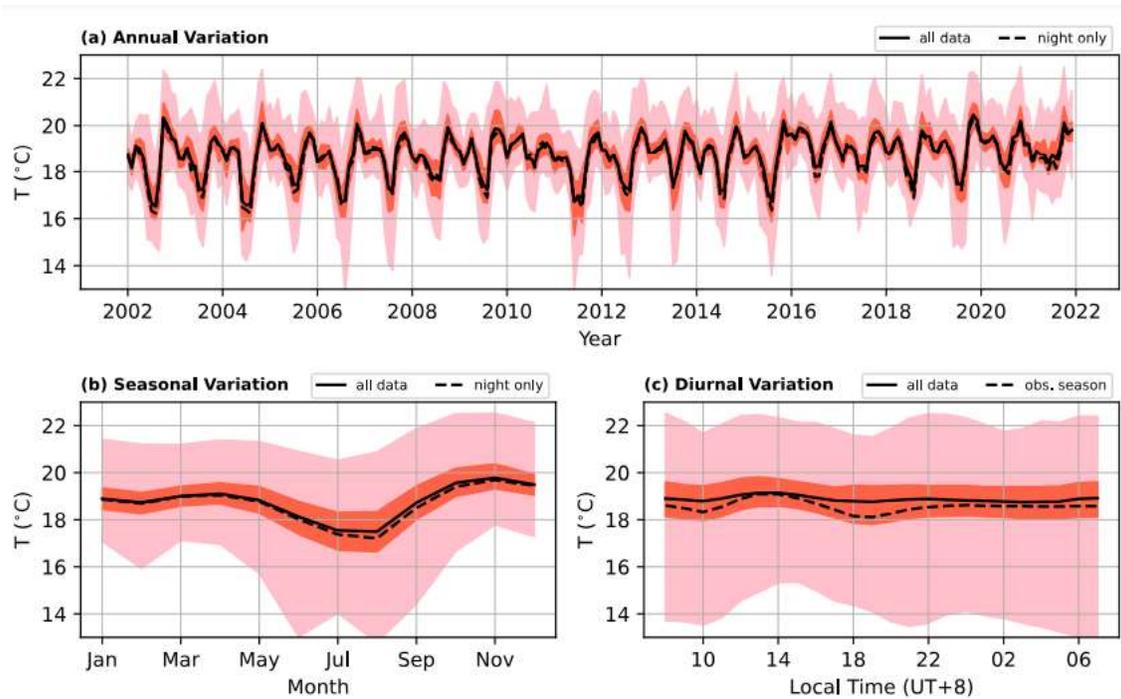

Figure 5. Same as Figure 3, but for horizontal surface temperature.

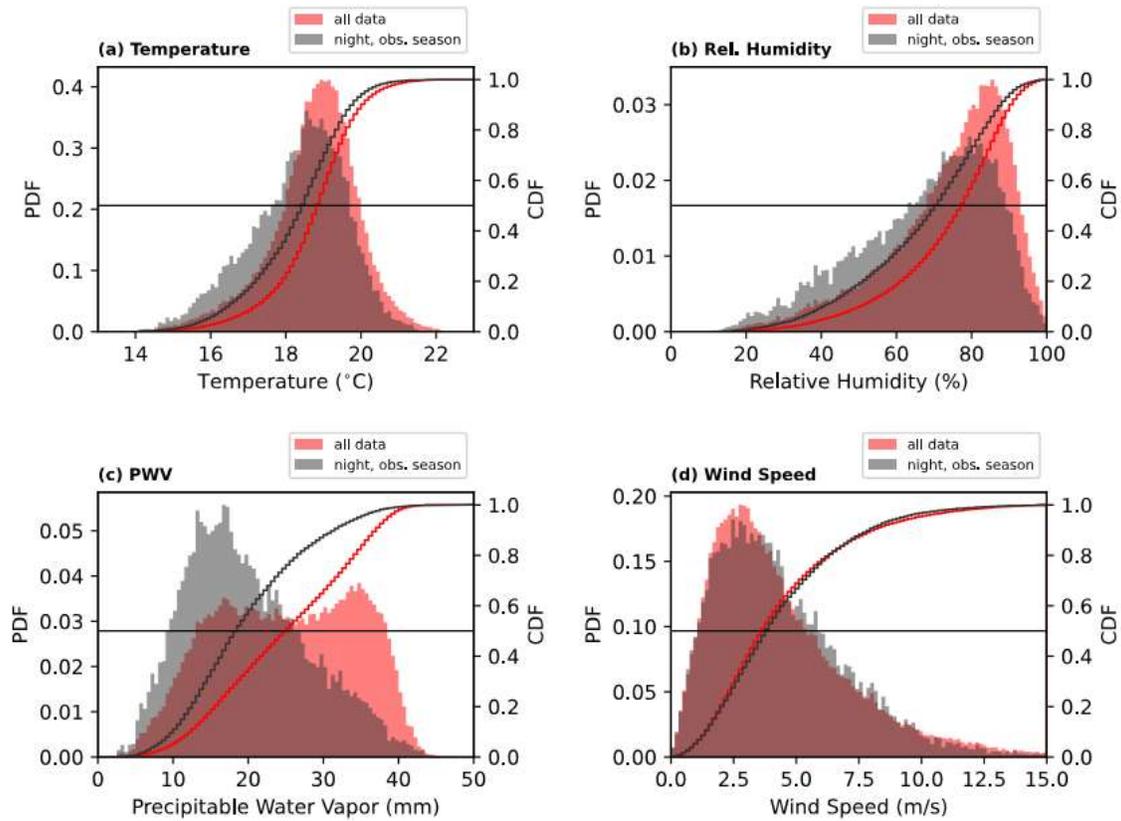

Figure 6. The histogram and cumulative frequency of meteorological parameters at Timau. The graphs in gray are overlaid to highlight the conditions during the observing season.

Disregarding the observed temporal variation, the overall distribution of the surface temperature at Timau is presented on the upper-left panel of Figure 6. It is clear that the distribution of the nighttime temperature deviates from the general distribution.

3.3 Wind Speed

According to the analyzed ERA5 data, 95% of the wind speed is below 10 m/s (see Figure 6). The overall distribution is characterized by a median speed of 3.9 m/s and IQR of 3.4 m/s. The same distribution is observed for the subset of the data during the observing season. Erratic variation of the horizontal wind speed is exhibited in the upper panel of Figure 7 even though the cyclical pattern can be seen. The seasonal plot shows a noticeable variation in wind speed, in which the highest occurs in January with a median value of about 6.0 m/s. This peak is preceded by an annual minimum of 3.5 m/s in December. Based on the aggregated data, the secondary maxima of 4.2 m/s is reached in November.

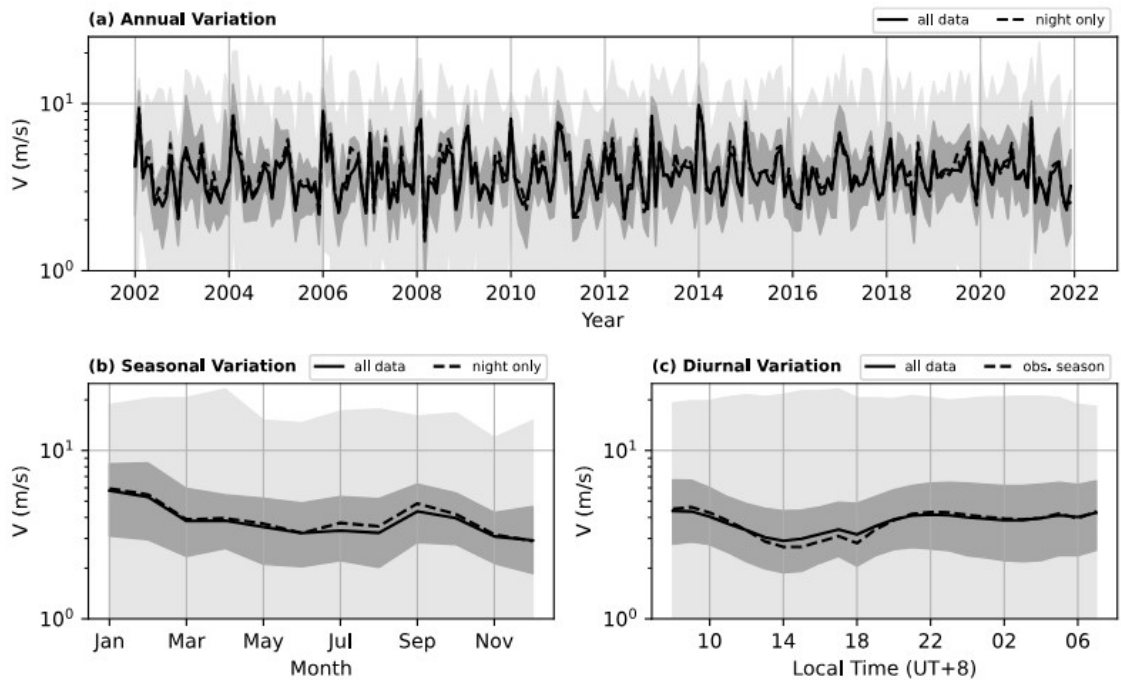

Figure 7. Same as Figure 3, but for horizontal wind speed. A logarithmic scale is used for the vertical axis.

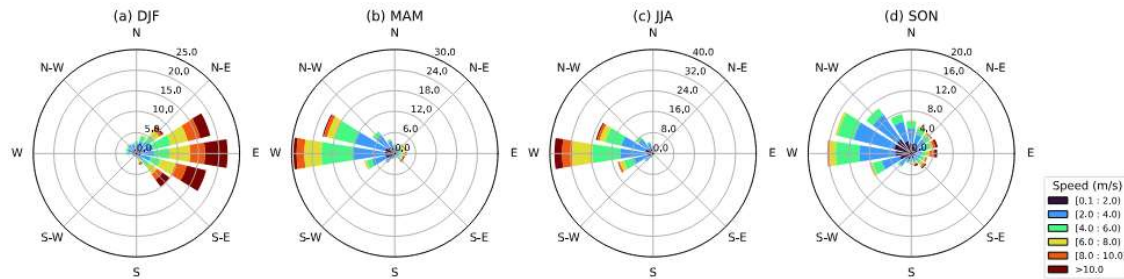

Figure 8. Windrose diagram for each quarter displays the distribution of wind speed and direction at Timau.

The diurnal variation of horizontal wind speed shows calm conditions with lower wind speeds normally occurring between 12.00 to 18.00 local time. When the night falls, the wind starts to get stronger and then stays at around 4.0 m/s overnight. However, the hourly aggregate of nighttime wind speeds has a wider spread as implied by IQR that is 100% larger than the IQR during the day.

From the windrose diagrams in Figure 8, the first quarter (December-January-February) experiences most of the strong winds that are directed eastward. Statistically, the chance of experiencing strong (> 10 m/s) winds is approximately 15%. Meanwhile, westward winds with lower speeds become more dominant during the other quarters. This is consistent with the results from Alifdini, Shimada, and Wirasatriya 2021 that a part of the Australian winter monsoons blows mainly through the Arafura Sea to the west along the southern Indonesian islands before reaching the Southern China Sea. This stream is articulated as westward winds over Timor Island.

The wind is an important factor in the operation of an observatory because it determines the thermal flushing of the observatory itself. On the other hand, strong winds can cause vibrations to the telescope or turbulence in the atmosphere, leading to distorted or blurry images of celestial objects. Setting a maximum limit of wind speed for day-to-day operation is also essential to ensure the safety of the telescope, instruments, and operators. From a different perspective, further site and facility development should consider the strong winds that may occur occasionally on site. As displayed on the lower right panel of Figure 6, the wind speed at Timau follows a Weibull distribution function (T and B 1984), with a chance of experiencing > 10 m/s wind is approximately 5%.

3.4 Humidity

Low humidity conditions are preferred for astronomical observations. The humidity surrounding an observatory influences the performance of the telescopes and instruments. For instance, humidity leads to the formation of dew on the telescope mirrors or in the instruments. This should be avoided by any means, including an interruption/termination of the observation sequence when the relative humidity exceeds 90%. However, relatively high humidity in Timau is an inevitable condition.

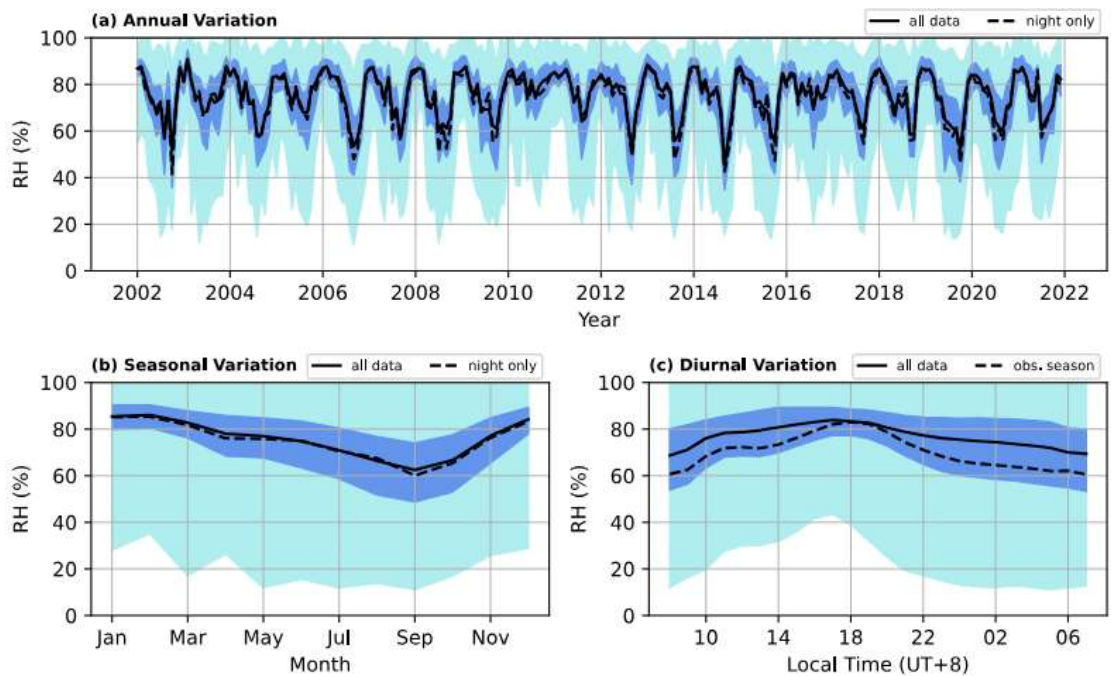

Figure 9. Same as Figure 3, but for relative humidity.

As indicated in Figure 6, the relative humidity at Timau is skewed to the left with an overall median of 77.7%. During the night in April-October, the median humidity is lowered to 70.9%. Fortunately, the chance of getting > 90% relative humidity is around 7% over this period.

Temporal variation of the humidity at Timau can be observed in Figure 9. In terms of monthly median, the relative humidity at this location varies from around 50% to about 90%. From the annual plot, 2010 and 2016 are considered anomalous years with relatively high humidity levels throughout the year. The lowest humidity is normally achieved in September while the peak is reached in January when the precipitation is also at its maximum. How the humidity varies over the year is almost similar to the variation of the surface temperature, but a time lag of about 1 month is present.

Due to the accumulation of evaporation during the day, the relative humidity increases to the daily maximum at around 18.00 local time. It then declines to the driest condition at around 07.00 local time. Interestingly, the daily minimum of humidity during the observing season is lower than the one computed from the whole data.

3.5 Precipitable Water Vapor

In line with relative humidity, precipitable water vapour (PWV) plays a crucial role in astronomical observations. It can affect atmospheric transparency, especially in the infrared and submillimeter bands. The overall distribution of PWV at Timau is presented in Figure 6 while the variations of this meteorological parameter at different temporal scales are displayed in Figure 10.

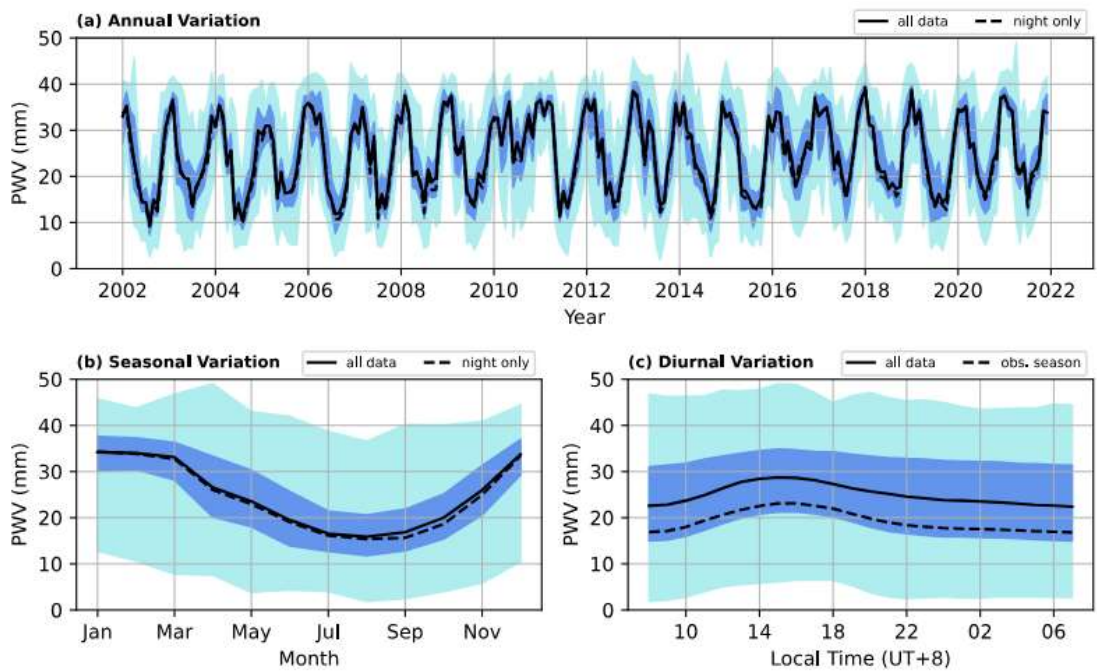

Figure 10. Same as Figure 3, but for precipitable water vapour.

At Timau, PWV fluctuates between 3 to 45 mm. The overall distribution of PWV has dual peaks that reflect the two main seasons in the region. During the observing season that takes place from April to October, the PWV tends to be lower than the other time. During this season, the median value of nighttime PWV is 18.5 mm, significantly lower than the global median of 25.1 mm. Qualitatively, this level of PWV (≥ 10 mm) indicates an extremely poor condition for infrared observation (e.g., Kidger et al. 1998). However, this can still be tolerated for near-infrared and optical observations. Based on the same subset (nighttime and observing season), the proportion of night with PWV lower than 10 mm is only 9.7%.

In a year, the lowest value of the PWV occurs in August with a median value of 17 mm, whereas the highest occurs in January with a median value of 35 mm. The distribution of PWV throughout the year seems to be invariant as indicated by nearly homogeneous monthly IQR. The diurnal variation of the PWV exhibits a small bump around 15.00 local time, while the lowest value of this parameter occurs at 08.00 local time. During the observing season, the hourly median PWV deviates from the global median by 5 mm. Flat PWV under 20 mm is retained after 20.00 local time. Consequently, optimum transparency at near-infrared is achieved for more than 80% of the nighttime.

3.6 Usable Nights

To further contextualize the results above, we estimate the percentage of usable nights at Timau according to the ERA5 dataset. Following the common practices, such as in Ningombam et al. 2021, we define usable night as the night (between 18.00 and 06.00 local time) when the cloud cover is under 0.375 (3 oktas) for at least 4 hours without interruption. This criteria also defines spectroscopic night. Better than that, photometric conditions can be achieved if the cloud cover is less than 0.125 for 4 hours or more. During these conditions, the relative humidity shall not exceed 90%. Figure 11 presents the monthly percentage of usable nights that vary from around 15% in January to the maximum of 80% in August. During the observing season, there are 68% usable nights, and 47% photometric nights. The yearly average of the percentage of usable nights is ∼ 50%, which is lower than quoted values (∼ 66%) from previous studies (Hidayat et al. 2012; Priyatikanto et al. 2023). Interestingly, overestimation of cloud cover or underestimation of clear sky fraction based on the ERA5 dataset was reported in literature (e.g., Lei et al. 2020).

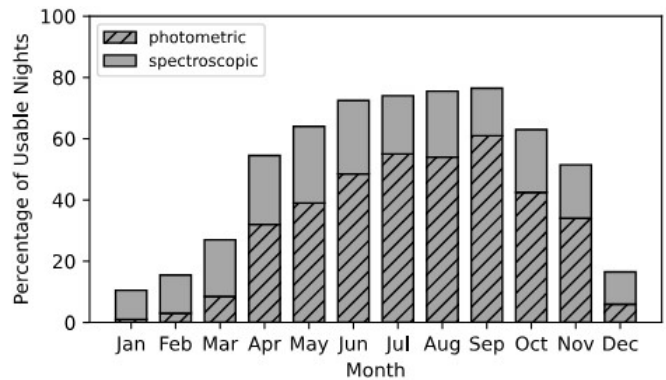

Figure 11. Monthly percentage of usable nights categorized as photometric and spectroscopic nights.

The observing season at Timau lasts from April to October during which little or no rainfall and minimum cloud cover are recorded. However, this season is also the most vulnerable season for the occurrence of wildfires due to the fire-prone conditions of the land. Currently, the Local Disaster Management Agency categorizes the area surrounding Timau as having a medium-high risk of wildfires . Direct and indirect consequences can be threatening to the observatory. In the case of a wildfire, the particles released by this can disturb astronomical observations which are sensitive to airborne particles. Outside Timor Island, the threat of air-polluting wildfires may come from Australia, the western parts of Indonesia (especially

Kalimantan and Sumatra), and southern Papua (Reddy and Sarika 2022). Fortunately, the seasonal winds keep Timau away from this threat. Australian wildfires may occur in summer time during which Timau experiences the wet season with monsoonal winds from the west. Conversely, the Kalimantan and Sumatra wildfires occasionally happen during the dry season when the winds blow westward. Despite its unlikeliness, the smoke from southern Papua can be a disturbance if the south-westerly winds dominate in the dry season.

3.7 Extreme Conditions

Among the meteorological parameters studied, wind speed has a right-skewed distribution where the extreme values on the right require attention. During extreme conditions with very strong wind speed, the operation at the observatory should be paused. To ensure safety, telescope enclosures and other supporting facilities should withstand the strong winds up to a certain level. Based on the results presented in Section 3.3, the distribution of wind speed at Timau has a 95 percentile of 10.1 m/s while the maximum value according to the 2002- 2021 ERA5 dataset reaches 23 m/s. Beyond this value, the occurrence of tropical cyclones should be taken into account in the context of future planning and development. Seroja cyclone which took place in April 2021 becomes an example. With the peak inland activity of two days, this cyclone partially damaged some facilities at Timau, including the outer part of the main telescope building.

Timor Island is located close to the Australian tropical cyclone basin (see Figure 1). Although the landfall of a tropical cyclone rarely happens on this island or in the region of Indonesia in general (Mulyana et al. 2018), the occurrence of these extreme phenomena should be mitigated. Based on multiyear records of tropical cyclones, some studies (Chand et al. 2019) reached some important conclusions. Firstly, tropical cyclones mostly emerge at the cyclogenesis region around the latitude of 11° S and dissipate around the latitude of 21° S. Some cyclones like Seroja were generated in the Indonesian maritime continent. Strong wind and enhanced precipitation become the direct effects of the cyclone (Mulyana et al. 2018; Putri et al. 2023). Secondly, the period between November to April is regarded as the cyclone season in the region. This period partially overlaps the observing season at Timau. Thirdly, based on the record spanning from 1960 to 2009, the estimated rate of Australian tropical cyclones is about 12 cyclones per year. This rate tends to decrease over time, but La Niña may elevate the rate of occurrence. Lastly, the majority of cyclones happen in the western part of the Australian cyclone basin. Cyclones with category 3 and wind speeds of 35 – 45 m/s are common in this region. Meanwhile, the occurrence and mean intensity of the cyclones in the central basin (south of Timor Island) are significantly lower compared to the values associated with the western basin. Considering these key points, it is reasonable to set the maximum wind speed at Timau as 30 m/s. This reference value is important for the further development of the Timau site. For instance, a 30-m class radio telescope is listed on the masterplan for Timau (Mumpuni et al. 2018; Huda et al. 2021), Thus, this huge structure should withstand extreme conditions, including rarely occurring tropical cyclones.

## 4. DISCUSSION
4.1 Comparison with Other Sites

As a new observatory site, the site of Timau Observatory should be characterized thoroughly to get some glimpses into the potency, advantages, and limitations of this site. An interesting benchmark was provided by Hellemeier and Hickson 2019 who investigated cloud cover, precipitable water vapour, wind speed at 200 hPa, vertical wind velocity, and aerosol index, for some astronomical sites worldwide using reanalysis datasets. Extended with some other sites, Table 1 summarizes the weather characteristics of the sites. The results showed that those sites are mostly dry with PWV below 10 mm and away from the equatorial band. Ideal sites where very large telescopes are placed, like Paranal and Mauna Kea, are unequal comparisons to

Timau. However, Timau's characteristics can be considered moderately good and comparable to the other sites located within or close to the tropical band and the sites with relatively low elevations.

Ascension Island, situated at the same latitude and geographical characteristics as Timau Observatory has PWV of 7.9±0.7 and humidity of around 69%. Taking advantage of its geographic position, Ascension Island hosts a 1.3-m telescope for orbital debris observations (Lederer, Buckalew, and Hickson 2019). In tropical regions, the PWV is heavily influenced by monsoons. Besides that, site altitude is a primary factor influencing yearly PWV averages, the atmosphere over higher sites contains less water vapour on average.

Siding Spring Observatory in Australia is not overly dry from the perspective of astronomy. Abbot et al. 2021 reported the median humidity of about 60% based on a decade-long archive of weather data while Hellemeier and Hickson 2019 quoted PWV = 6.7 ± 0.7 mm according to ERA-40 dataset. The median temperature at that site is about 14 ºC while the range between the daily minimum and maximum is about 8 ºC. In terms of wind speed, Timau and Siding Spring have similar characteristics, e.g., having median wind speed around 3.5 m/s and IQR of 3.3 m/s. Between 2010 and 2020, the maximum recorded wind speed at Siding Spring was around 40 m/s (average from different sensors), which is higher than the expected extreme value at Timau. Regarding cloud cover, Siding Spring Observatory has an observing season that coincides with Timau's.

The Xinglong Observatory in China can be another fair comparison. According to Zhang et al. 2015, this observatory has 55% humid time during the summer due to monsoon and rainfall. However, the humidity and clear sky fraction are relatively good during the winter. During this season, the median wind speed is under 2.0 m/s, lower than Timau's median wind speed. Aside from this relatively calm condition, the median seeing at Xinglong is about 2.0 arcseconds.

While the climate at Timau may not be the best for astronomical observations compared to some existing observatories, its geographic location offers a distinct advantage. Southeast Asia has a scarcity of major observatories. Timau's position can bridge the geographical gap between facilities in the Pacific, Australia, and East Asia, providing a more complete picture of the night sky. Furthermore, a 4-meter class telescope at Timau would be particularly valuable for following up on transient phenomena, where ideal atmospheric conditions are less critical.

4.2 Future Projection

A new observatory is expected to work productively and sustainably as long as possible, or at least for the next 50 years. Based on the past, present, and projected characteristics, We can anticipate how Timau National Observatory should be managed and operated.

Haslebacher et al. 2022 investigated trends in observing conditions for eight major astronomical sites based on the climate projection models until 2050. The parameters they studied were temperature, specific humidity, PWV, cloud cover, and astronomical seeing. They found positive trends in temperature and precipitable water vapour while the humidity tended to decrease. The warming climate also affect Timau for the next decades. Assuming the intermediate scenario of Representative Concentration Pathway (RCP) 4.5, the average temperature at the Timau area (0 masl) is projected to increase by 1.7ºC.

While climate change is projected to increase wildfire risk globally, with hotter, drier conditions (Herawati and Santoso 2011) particularly impacting higher latitudes (Jolly et al. 2015; Senande-Rivera, Insua-Costa, and Miguez-Macho 2022), tropical regions may experience a lesser increase. However, this doesn't eliminate the threat entirely. Wildfires could significantly disrupt observations at the observatory as the primary observing season coincides with the region's dry season.

Table 1. Summary of weather parameters from astronomical sites around the world.

| Site | Geo-coord (lat, long) | Elev. (m) | CC (%) | Temp. (°C) | Wind (m/s) | RH (%) | PWV (mm) | Ref. |
|---|---|---|---|---|---|---|---|---|
| Timau | (9.58S, 123.95E) | 1300 | 0.61[a] | 18.9[a] | 3.70[a] | 77.7[a] | 25.1[a] | This paper |
| Ascension | (7.93N, 14.42W) | 150 | 0.53[a] | 26.0 | | | 7.91[a] | Hellemeier and Hickson 2019 |
| Devasthal | (29.37N, 79.68E) | 2450 | 0.40[a] | | | | 11.22[a] | Hellemeier and Hickson 2019; Radu et al. 2012 |
| Eastern Antatolia | (39.78N, 41.23E) | 3100 | 0.47[a] | | | | 5.18[a] | Sarazin 1990 |
| La Silla | (28.25S, 70.73W) | 2400 | | | 1.5 | 37.0 | 3.90 | Hellemeier and Hickson 2019 |
| Mauna Kea | (19.82N, 155.47W) | 4200 | 0.25[a] | 2.29[a] | 10.0 | 25.0 | 2.22[a] | Hellemeier and Hickson 2019 |
| Paranal | (24.63S, 70.40W) | 2635 | 0.08[a] | 12.8[a] | 2.2 | 14.5 | 2.70[a] | Hellemeier and Hickson 2019; Lombardi et al. 2009 |
| Roque de los Muchachos | (28.77N, 17.88E) | 2395 | 0.17[a] | 9.8[a] | 5.0[a] | 22.5[a] | 3.90 | Garcia-Lorenzo et al. 2010; Hidalgo et al. 2021; Hellemeier and Hickson 2019 |
| San Pedro Martir | (31.04N, 115.46W) | 2800 | 0.31[a] | 8.7 | 5.8 | 28.5 | 2.63 | Tapia et al. 2007 |
| Siding Spring | (31.27S, 149.06E) | 1165 | 0.40[a] | 14.1 | 3.4 | 60.3 | 6.73[a] | Abbot et al. 2021; Hellemeier and Hickson 2019 |
| Sutherland | (32.38S, 20.81W) | 1800 | 0.27[a] | | | | 9.65[a] | Hellemeier and Hickson 2019 |
| Xinglong | (40.39N, 117.58E) | 900 | 0.37[a] | 7.8 | 1.65 | 60.0 | | Zhang et al. 2015 |

[a]based on atmospheric reanalysis dataset

Table 2. Statistical summary of the meteorological parameters at Timau. The summary of the parameters during the observing season is presented in the rightmost column.

| Cloud Cover | | | | | | | | | | | | | | |
|---|---|---|---|---|---|---|---|---|---|---|---|---|---|---|
| P25 | 0.81 | 0.66 | 0.50 | 0.18 | 0.12 | 0.08 | 0.06 | 0.03 | 0.04 | 0.09 | 0.18 | 0.65 | 0.17 | 0.07 |
| median | 0.97 | 0.96 | 0.91 | 0.56 | 0.41 | 0.29 | 0.28 | 0.23 | 0.30 | 0.41 | 0.51 | 0.95 | 0.61 | 0.35 |
| P75 | 1.00 | 1.00 | 0.99 | 0.94 | 0.80 | 0.73 | 0.75 | 0.68 | 0.73 | 0.82 | 0.90 | 1.00 | 0.95 | 0.79 |
| P95 | 1.00 | 1.00 | 1.00 | 1.00 | 0.99 | 0.98 | 0.97 | 0.98 | 0.98 | 0.99 | 1.00 | 1.00 | 1.00 | 0.99 |
| Temperature (°C) | | | | | | | | | | | | | | |
| P25 | 18.4 | 18.3 | 18.6 | 18.7 | 18.3 | 17.4 | 16.7 | 16.6 | 18.0 | 19.0 | 19.3 | 19.0 | 18.2 | 17.5 |
| median | 18.9 | 18.7 | 19.0 | 19.1 | 18.8 | 18.1 | 17.5 | 17.5 | 18.7 | 19.6 | 19.8 | 19.5 | 18.9 | 18.5 |
| P75 | 19.3 | 19.2 | 19.4 | 19.6 | 19.4 | 18.8 | 18.3 | 18.3 | 19.4 | 20.2 | 20.4 | 19.9 | 19.5 | 19.2 |
| P95 | 20.0 | 19.8 | 20.1 | 20.2 | 20.1 | 19.7 | 19.2 | 19.4 | 20.4 | 21.2 | 21.3 | 20.8 | 20.5 | 20.1 |
| Wind speed (m/s) | | | | | | | | | | | | | | |
| P25 | 3.10 | 2.96 | 2.35 | 2.62 | 2.11 | 2.05 | 2.22 | 2.02 | 2.85 | 2.77 | 2.14 | 1.86 | 2.33 | 2.41 |
| median | 5.78 | 5.30 | 3.81 | 3.84 | 3.51 | 3.23 | 3.35 | 3.24 | 4.34 | 3.96 | 3.09 | 2.93 | 3.70 | 3.87 |
| P75 | 8.30 | 8.43 | 5.94 | 5.44 | 5.20 | 4.86 | 5.31 | 5.17 | 6.30 | 5.59 | 4.27 | 4.63 | 5.76 | 5.87 |
| P95 | 12.48 | 13.39 | 10.27 | 8.73 | 8.51 | 8.36 | 9.09 | 10.04 | 9.61 | 8.98 | 6.18 | 9.36 | 10.10 | 9.39 |
| Relative humidity (%) | | | | | | | | | | | | | | |
| P25 | 80.0 | 80.5 | 76.3 | 68.5 | 67.8 | 63.5 | 58.7 | 51.9 | 48.8 | 53.1 | 65.6 | 78.1 | 65.5 | 55.6 |
| median | 85.5 | 86.1 | 82.8 | 78.0 | 77.0 | 74.8 | 70.9 | 66.4 | 62.5 | 66.7 | 77.1 | 84.2 | 77.7 | 71.4 |
| P75 | 90.3 | 90.4 | 87.8 | 85.8 | 84.7 | 83.4 | 80.3 | 76.8 | 74.0 | 77.5 | 84.8 | 89.3 | 85.8 | 81.5 |
| P95 | 96.3 | 96.3 | 94.2 | 93.7 | 92.6 | 92.2 | 90.7 | 88.7 | 88.3 | 89.4 | 92.0 | 95.0 | 93.6 | 91.3 |
| Precipitable water vapor (mm) | | | | | | | | | | | | | | |
| P25 | 30.1 | 30.5 | 28.2 | 20.1 | 18.1 | 13.9 | 12.7 | 11.7 | 12.8 | 15.4 | 20.6 | 29.2 | 17.3 | 13.6 |

| | | | | | | | | | | | | | |
|---|---|---|---|---|---|---|---|---|---|---|---|---|---|
| median | 34.2 | 34.0 | 33.1 | 26.5 | 23.6 | 19.5 | 16.5 | 15.9 | 16.9 | 20.0 | 26.0 | 33.8 | 25.1 | 18.5 |
| P75 | 37.6 | 37.3 | 36.3 | 33.3 | 30.3 | 25.7 | 21.4 | 20.6 | 21.8 | 25.1 | 31.3 | 36.9 | 33.0 | 25.1 |
| P95 | 40.9 | 40.3 | 39.3 | 39.5 | 37.6 | 33.4 | 31.9 | 27.5 | 30.1 | 33.4 | 37.2 | 40.2 | 38.8 | 35.0 |

## 5. CONCLUSION

As a new scientific facility, the astroclimate characteristics of Timau National Observatory should be understood thoroughly. This paper analyses important meteorological parameters from ERA5 2002-2021, including surface temperature, horizontal wind speed, humidity, and precipitable water vapour (PWV). In addition to the quantities presented in Table 2, the following points summarize our results:

- According to the ERA5 dataset, there is approximately 50% of usable nights in a year. A significant discrepancy (~ 15%) exists between this estimate and the satellitebased estimate from Hidayat et al. 2012 and Priyatikanto et al. 2023.
- Temperature at Timau varies between 18 ºC to 20 ºC with the median of 18.9 ºC. During the observing season that lasts from April to October, the temperature is marginally lower. In line with the need for the astronomical observatory, the daily fluctuation of temperature on site is maintained low. In aggregate, the difference between the daily maximum and minimum is around 1.5 ºC.
- The distribution of wind speed at Timau follows the Weibull distribution with a median of 3.64 m/s. Asian and Australian monsoons influence the directional pattern of the wind during the wet and dry seasons, respectively. Timau may experience strong wind (up to 30 m/s) close to the Australian tropical cyclone basin, though the chance is low.
- High humidity and water content above Timau become the weakest aspect of this site. During the observing season, the odds of getting > 90% humidity at night is about 7%. In the same period, the median PWV is 18.5 mm, categorized as extremely poor for infrared observations. However, photometric observations at near-infrared are still promising. Compared to an ideal site with PWV = 1 mm, Timau has fair atmospheric transparency.

While the main telescope is planned to be completely installed in 2024, the above-mentioned weather characteristics only partially capture the conditions at Timau. Additional parameters are required to ensure the operation and further development of this new observatory goes well. Among others, seeing estimations based on the ERA5 dataset and on-site measurements are currently ongoing.

After commencing the operation of the observatory, we do hope that there will be new astronomical findings and insights so this facility will act as the gate to the new era of Indonesian astronomy. In the long run, hopefully, this facility will also reduce the disparity between the eastern and the western parts of Indonesia. The growing interest in science and technology related to the science of astronomy will raise the educational level in the province of East Nusa Tenggara, so the quality of the local human resources will increase and so will the quality of life in the area.

## 6. ACKNOWLEDGEMENT

This research is partially funded by Organisasi Riset Nanoteknologi dan Material (ORNM-BRIN). We thank the ECMWF for providing hourly meteorological data.